\documentclass [preprint]{aastex}

\providecommand{\h}{}
\providecommand{\vih}{}
\providecommand{\v}{}
\renewcommand{\v}{{\it V} }
\renewcommand{\h}{{\it H} }
\renewcommand{\i}{{\it I} }
\renewcommand{\vih}{{\it VIH} }

\begin{document}

\title{Multiwavelength Observations of 
the Black Hole Candidate XTE J1550-564 during the 2000 Outburst}

\author{Raj K. Jain\altaffilmark{1,2}, Charles D. Bailyn\altaffilmark{2},
Jerome A. Orosz\altaffilmark{3}, Jeffrey E. McClintock\altaffilmark{4}, and
Ronald A. Remillard\altaffilmark{5}}

\altaffiltext{1}{Department of Physics, Yale University,
 P.\ O.\ Box 208120, New Haven, CT 06520-8120, raj.jain@yale.edu} 
\altaffiltext{2}{Department of Astronomy, Yale
 University, P.\ O.\ Box 208101, New Haven, CT 06520-8101,
 bailyn@astro.yale.edu}
\altaffiltext{3}{Sterrenkundig Instituut, Universiteit Utrecht, 
Postbus 80.000, 3508 TA Utrecht, The Netherlands, 
 J.A.Orosz@astro.uu.nl}
\altaffiltext{4}{Harvard-Smithsonian 
Center for Astrophysics, 60 Garden Street, Cambridge, MA 02138-1516, 
jem@cfa.harvard.edu}
\altaffiltext{5}{Center for Space Research,
Massachusetts Institute of Technology, Cambridge, MA 02139-4307, 
rr@space.mit.edu}

\begin{abstract}

We report optical, infrared, and X-ray light curves for the outburst, in 2000,
of the black hole candidate XTE J1550-564.  We find that the start of the
outburst in the $H$ and $V$ bands precedes that seen in the RXTE All Sky
Monitor by $11.5\pm 0.9$ and $8.8 \pm 0.6$ days, respectively; a similar delay
has been observed in two other systems.  About 50 days after the primary
maxima in the \vih light curves, we find secondary maxima, most prominently in
\h.  This secondary peak is absent in the X-ray light curve, but coincides with
a transition to the low/hard state.  We suggest that this secondary peak may
be due to non-thermal emission associated with the formation of a jet.
\end{abstract}

\keywords{black hole physics --- X-rays: stars --- stars: individual
(XTE J1550-564)}

\section{Introduction}

X-ray novae (XNe) are mass transferring binaries in which long periods of
quiescence (when the X-ray luminosity is $\le 10^{33}$ergs s$^{-1}$) are
occasionally interrupted by luminous X-ray and optical outbursts \citep{ts96}.
XNe provide compelling evidence for the existence of stellar mass black holes
\citep{cowley,van95}, since they can be shown to contain compact objects whose
mass exceed the maximum stable limit of a neutron star, which is $\approx
3M_{\sun}$ \citep{ch96}.  Observations of the companion star in quiescence can
lead to a full understanding of the orbital parameters of the system,
including the masses of the binary components and the orbital inclination
\citep{bai98}.

A detailed understanding of the accretion flow in these objects is of
considerable importance, since the behavior of the flow close to the event
horizon may give rise to tests of general relativity in the strong field
limits \citep{mcclintock98}.  During their outburst cycles, XNe generally
display the complete range of spectral states, from quiescent (``off'') to
``low/hard'' to ``high/soft'' to ``very high''.  They therefore present unique
opportunities to study all of these kinds of accretion flows in a situation in
which the geometry of the binary system is well understood.  Multiwavelength
observations covering the entire outburst cycle are of particular importance,
since the different wavelength regimes probe different aspects of the
accretion flow.  The RXTE satellite, with its All Sky Monitor (ASM; Levine et
al. 1996) and unprecedented scheduling flexibility, has set a new standard in
X-ray coverage of XNe outbursts.  By contrast, optical-IR (OIR) coverage over
an entire outburst is often hard to arrange from ground-based facilities.
Some of the difficulties are inherent to Earth-based observing (e.g. weather
and daylight) but it is also true that many facilities are scheduled in short
blocks determined well in advance, which precludes following unpredictable XN
outbursts which can last for several months.

The YALO Consortium (Yale, A.U.R.A., Lisbon University, and The Ohio State
University) was established in part to address the latter difficulty (Bailyn
et al. 2000).  YALO operates the Yale 1m telescope at Cerro Tololo
Interamerican Observatory.  Permanently mounted on the telescope is the
ANDICAM instrument, constructed at Ohio State University, which contains both
a TEK $2048\times 2048$ CCD and a Rockwell $1024\times 1024$ HgCdTe IR array.
The light path contains a dichroic which allows optical and near IR images to
be obtained simultaneously at the same position on the sky.  The telescope is
scheduled as a queue designed to optimize long term monitoring and quick
response to unexpected events.  Thus YALO is well-suited to providing
optical-infrared light curves of XNe outbursts.

We have used YALO to monitor the XN XTE J1550-564 since its discovery with the
RXTE ASM on 1998 September 6 (Smith et al.\ 1998).  Details of the initial
outburst (1998-1999) have been published by Sobczak et al. (1999, 2000a,
2000b), Jain et al. (1999, 2001; hereafter Paper I and II, respectively),
and Remillard et al. (1999).  After a a brief period of
quiescence between MJD 51346 to 51600, the source returned to outburst between
late March and July 2000 \citep{smith00,ms00,jb00,mws00}.  A re-flare was also
detected in the black hole candidate GRO J1655-40 \citep{orosz97}, a source
that is similar to XTE J1550-564 with respect to its outburst light curve
morphology, orbital period, and X-ray QPO behavior.  Recently (2001 January)
XTE 1550-564 started an unprecedented third outburst (Tomsick et al. 2001;
Jain, Bailyn \& Tomsick 2001).  Radio observations of XTE 1550-564 during the
second outburst in 2000 have been reported by Corbel et al. (2001).  In this
paper we report our optical and infrared light curves of that outburst --- our
observations provide a well-sampled light curve, which is particularly unusual
in the infrared.  In \S~\ref{obs} we discuss observations and data reductions
and in \S~3 compare our OIR light curves with the RXTE ASM X-ray light curve.
We discuss two important features of the light curves: the different times of
onset of brightening in the various bandpasses, and the presence of a
secondary peak in the OIR light curve, which is most striking in the infrared.
In \S~4 we list some of the questions prompted but currently unanswered by
this work.

\section{Observations and Data Reduction \label{obs}} 

We obtained \v and \h images of XTE 1550-564 every clear night between MJD
51594.28 (February 20, 2000) and 51809.00 (September 22, 2000).  We also
obtained \i band observations during the outburst between MJD 51645.16 (April
11) and MJD 51717.14 (June 22).  Exposure times were 1200s in \v and 600s to
900s in \i.  We obtained 345 frames in \v and 120 frames in \i.  \h-band
images were constructed by averaging 4 to 12 contiguous dithered frames of 80
to 95 seconds each.  The dithered images were flat-field corrected, and then
used to create a median sky image, which was subtracted from the individual
frames before they were shifted and co-added.  In this manner we constructed
425 combined \h images.  Typically one to four images, and at times up to
eight images were obtained per filter per night.

The optical light curve was generated from differential magnitudes with
respect to the four neighboring non-varying stars shown in Figure 2 of Paper
II.  \h-band data were calibrated using standard stars reported in
\citet{per98}.  OIR light curves and RXTE ASM and hardness ratio plots are
displayed in Figure~1, and prominent morphological features are indicated,
when appropriate.  

\placefigure{fig1}

\section{Discussion}

In this paper, we will focus on two features of Figure 1, namely the delay
between the start of the outburst in the OIR and in the X-ray, and the
secondary peak, which is most prominent in the H-band light curve.  Additional
details of the light curve and its implications are discussed by Jain (2001).

\subsection{Initial Rise}

Figure 2 shows the start of the outburst.  A linear rise is fit to the initial
data in $V$, $H$, and X-ray bands.  Following the procedures described by
\citet{orosz97} for GRO J1655-40, and Jain et al. (in preparation) for Aql
X-1, we define the start of the initial rise as the time when the fit to the
linear portion of the earliest rise intersects the quiescent level (see Fig.
2). By this definition, the \h outburst began on MJD $51623.7 \pm 0.8$,
relative to the quiescent brightness of $\h=16.2\pm 0.1$ mag and the \v-band
outburst began on MJD $51626.4\pm0.4$, relative to the quiescent level of
$\v=21.6 \pm 0.2$ mag.  The dominant error in the starting time of the
outburst is the quiescent flux level, which varies intrinsically by up to
$\sim 0.3$ magnitudes in both bands (Paper II).  The beginning of the
X-ray rise was at 51635.2$\pm0.4$, implying an X-ray delay of $8.8 \pm 0.6$
days relative to the rise in \v and $ 11.5 \pm 0.9$ days relative to the rise
in \h.

\placefigure{fig2}

This source is the third XN in which an OIR outburst has been observed to
precede an X-ray outburst.  GRO~J1655-40 (Orosz et al. 1997) and Aql X-1 (Jain
et al., in preparation) have shown very similar X-ray delays; a somewhat
shorter delay has also been reported for a different outburst of Aql X-1
\citep{sh98a} based on a more sparsely sampled data set.  These observed
delays are difficult to explain in the context of a standard thin disk model,
since any delay ought to be on the front-propagation timescale, which is of
order hours.  Hameury et al.  (1998) proposed that a composite disk+ADAF
(Advection Dominated Accretion Flow) scenario could explain the delay in GRO
J1655-40, since in this case the observed delay would represent the time
required for the inner edge of the disk to propagate through the ADAF cavity
and reach the central source.  In this case, the relevant timescale is the
viscous timescale, which is on the order of days.

This explanation of the time delay receives some support from the fact that
the delays seen in the three sources are all of approximately the same length,
despite their differing orbital periods.  Their periods vary by over a factor
of three (19 hours for Aql X-1; 37 hours for XTE 1550-56; 63 hours for GRO
J1655-40), and the timescale on which a disturbance would propagate from the
outer parts of the disk or the secondary to the compact object might be
expected to scale with the Roche-lobe radius of the primary.  By contrast, if
the inner edge of the accretion disk in quiescence is fixed at some specific
distance from the compact object, then the time required for it to propagate
inwards might plausibly be similar in all systems.  Observations of more such
time delays, particularly in systems which have recurrent outbursts, such as
Aql X-1, will clearly be important in understanding the initial phases of XN
outbursts.  It will also be important to obtain a more physically motivated
picture of what causes the transition from disk to ADAF flows.

\subsection{The Secondary Flare}
The maximum flux levels of the \v and \i light curves were followed initially
by a period of exponential decay (see Fig.~\ref{fig1}b,c). Similarly in the
X-ray, the flux decayed exponentially while the spectra softened (see
Fig.~\ref{fig1}d,e).  On the other hand, the \h-band light curve decays more
sharply after reaching its peak and exhibits an exponential decay only after
MJD 51660 (see Fig.~\ref{fig1}a).  The e-folding time during this primary
decay is progressively shorter from \h to $V$, and is extremely short in the
X-ray (see Table~\ref{tab:lc}).

While the ASM light curve decays smoothly to quiescence, the $VIH$ light
curves are punctuated by a prominent secondary maximum (see
Fig.~\ref{fig1}a,b,c).  During this secondary flare, the \vih light curves
increased in brightness until MJD $51699 \sim 51705$, after which the source
brightness again decayed exponentially.  Compared to the peak brightness
during the initial rise, the \h band secondary flare peak is only $0.52 \pm
0.07$ mags fainter, relative to the initial peak, while the \i and \v
secondary flare peaks are $0.73 \pm 0.03$ and $0.99 \pm 0.01$ mags fainter,
respectively, than the initial peak values (see Fig.~\ref{fig2}).  To quantify
the amplitude, timing and width of the secondary peak, we parameterize the
decay light curves by fitting a Gaussian and an exponential decay to the $VIH$
light curves between MJD 51658 and 51741. Although the fits are statistically
poor, due to the obvious limitation of such a simple model, they are visually
compelling (see Fig.~\ref{fig3}). We find that the secondary peak is
significantly sharper in $H$, compared to the peaks in \i and \v which
occurred $3.6\pm 0.2$ and $ 4.8 \pm 0.2$ days after the \h band peak,
respectively.  Table 1 reports the parameters of our best fits to the Gaussian
plus exponential.

\placefigure{fig3}

In order to explain this secondary peak by X-ray reprocessing, one would have
to postulate that X-rays with energies above the ASM bandpass behaved in a way
that was not reflected at all in the ASM data.  This objection is perhaps not
insuperable given the change in X-ray hardness that occurred at about the same
time, when the source entered the low/hard state (see Fig.~\ref{fig1}e).  But
it appears unlikely to us that the secondary peak is reprocessed X-rays or any
other thermal emission.  A thermal origin for the excess flux would require a
low temperature to account for the dominance of the $H$-band, and therefore a
large volume to create the necessary flux.  Given the 1.5 day orbital period
(Paper II), such a large volume might be hard to accommodate.  Exact
calculations must await determinations of the geometry of the system which
will be carried out when the source finally reaches quiescence.  Instead, we
suggest that the secondary peak may be caused by non-thermal emission, perhaps
synchrotron radiation associated with a jet.  Jet emission may preferentially
occur when the source enters the low/hard state \citep{fender01}, which occurs
at $\approx 10\% $ of the Eddington limit \citep{emn97}.  This interpretation
is strengthened by the $\sim 1$mJy 8640 MHz radio detection on MJD 51697.14
\citep{corbel01}. Based on $V$, \i and radio data, \citet{corbel01} find an
inverted radio spectrum with a spectral index of $0.37\pm 0.1$, which has been
interpreted as evidence for optically thick synchrotron emission arising from
a compact and conical jet \citep{hj88}.  In the future, it may be useful to
use IR light curves as a trigger for extensive radio observations to determine
if jets do indeed accompany such IR-dominated secondary peaks.

\section {Unanswered Questions}

Our work poses a number of interesting questions. Are all XN outbursts
preceded by a rise in OIR flux approximately a week prior to the detection of
significant X-ray flux?  If so, is this timescale set by the viscous
propagation time through the ADAF (\S 3.1)?  Do secondary peaks of the kind we
observed occur frequently in IR light curves of XN outbursts?  Are the
``plateaus'' occasionally observed in optical decay light curves (e.g. see
Paper II) evidence of secondary outbursts which would have been more clearly
seen in the IR?  Are such secondary outbursts indeed associated with jet
production?  Do they always coincide with a transition to a low/hard state, or
to a specific flux level in the X-ray?

To answer these questions it will be necessary to carry out full
multiwavelength monitoring of XN outbursts in radio, OIR and high-energy
regimes.  While such campaigns are hard to arrange, the physical and
astronomical importance of accretion flows onto black holes and neutron stars
warrants the effort.  

\acknowledgements

We thank the two YALO observers, David Gonzalez Huerta and Juan Espinoza, for
providing data in a timely manner.  We would like to thank S. Tourtellotte and
E. Terry for their assistance with data reduction and J. Yong, R. Winnick, \&
B. Roscherr for their assistance with computer facilities as well as useful
discussions.  We would like to thank S. Corbel for discussions on radio data
and M. Nowak, T. Maccarone, and P. Coppi for extensive discussions on the
interpretations of the light curves.  Financial support for this work was
provided by the National Science Foundation through grant AST 97-30774.

\clearpage

 \begin{deluxetable}{lllllllllll} 
\tabletypesize{\scriptsize}
\tablewidth{0pt}
\rotate
\tablecaption{Summary of Optical and X-ray Light Curves  \label{tab:lc}}

\tablehead{ \colhead{} & \multicolumn{3}{c}{Initial Rise} & \colhead{} &
	\multicolumn{2}{c}{Primary Decay} & \colhead{} & 
\multicolumn{3}{c}{Secondary
	Flare} \\ \cline{2-4} \cline{6-7} \cline{9-11} \\ \colhead{} &
	\colhead{Start} & \colhead{Slope} & \colhead{Peak} & \colhead{} &
	\colhead{Dates} &\colhead{$\tau_{e}$} & \colhead{}& \colhead{Dates} &
	\colhead{Peak\tablenotemark{b}} & \colhead{FWHM\tablenotemark{b}} \\
 \colhead{Filter} &
	\colhead{(MJD-51000)} & \colhead{(mag/d)} & \colhead{(MJD-51000,mag)} &
	\colhead{} & \colhead{(MJD-51000)} & \colhead{(d)} &\colhead{} &
	\colhead{(MJD-51000)} & \colhead{(MJD-51000)} & \colhead{(d)} 
	 }

\startdata 
\h & $623.7\pm0.8$ & $-0.132\pm0.003$ & 654.3, $13.36\pm0.05$ & &
654.2$\sim$688 & $46.2\pm1.4$\tablenotemark{a} & & 688$\sim$747.2 &
$700.7\pm0.1$ & $16.7\pm0.2$ \\

\i & ... & ... & 653.4, $15.92\pm0.02$ & & 653.4$\sim$688 &
$33.6\pm0.3$\tablenotemark{c} & & ... & $704.3\pm0.2$ & $21.5\pm0.4$ \\

\v & $626.4\pm0.4$ & $-0.511\pm0.009$ & $653.4$, $17.89\pm0.02$ & &
$653.4\sim688$ & $30.4\pm0.3$\tablenotemark{d} & & $688\sim750.9$ &
$705.5\pm0.2$ & $19.4\pm0.4$ \\

ASM & $635.2\pm0.4$ & $0.0142\pm0.001$\tablenotemark{e} & 661.8,
$0.99\pm0.02$\tablenotemark{f} & & $661.8\sim701$ &
$10.8\pm0.2$\tablenotemark{g} & & ... & ... & ... \\

\enddata
\tablenotetext{a}{Fit to MJD 51660-51686}
\tablenotetext{b}{Determined by a Gaussian+linear fit to data between MJD
51658-51741}
\tablenotetext{c}{Fit to MJD 51655-51686}
\tablenotetext{d}{Fit to MJD 51655-51686}
\tablenotetext{e}{Crabs day$^{-1}$}
\tablenotetext{f}{Crabs}
\tablenotetext{g}{Fit to MJD 51661$\sim$51680}
\end{deluxetable}

\clearpage

\begin{center}
{\bf Figure Legends}
\end{center}


\figcaption[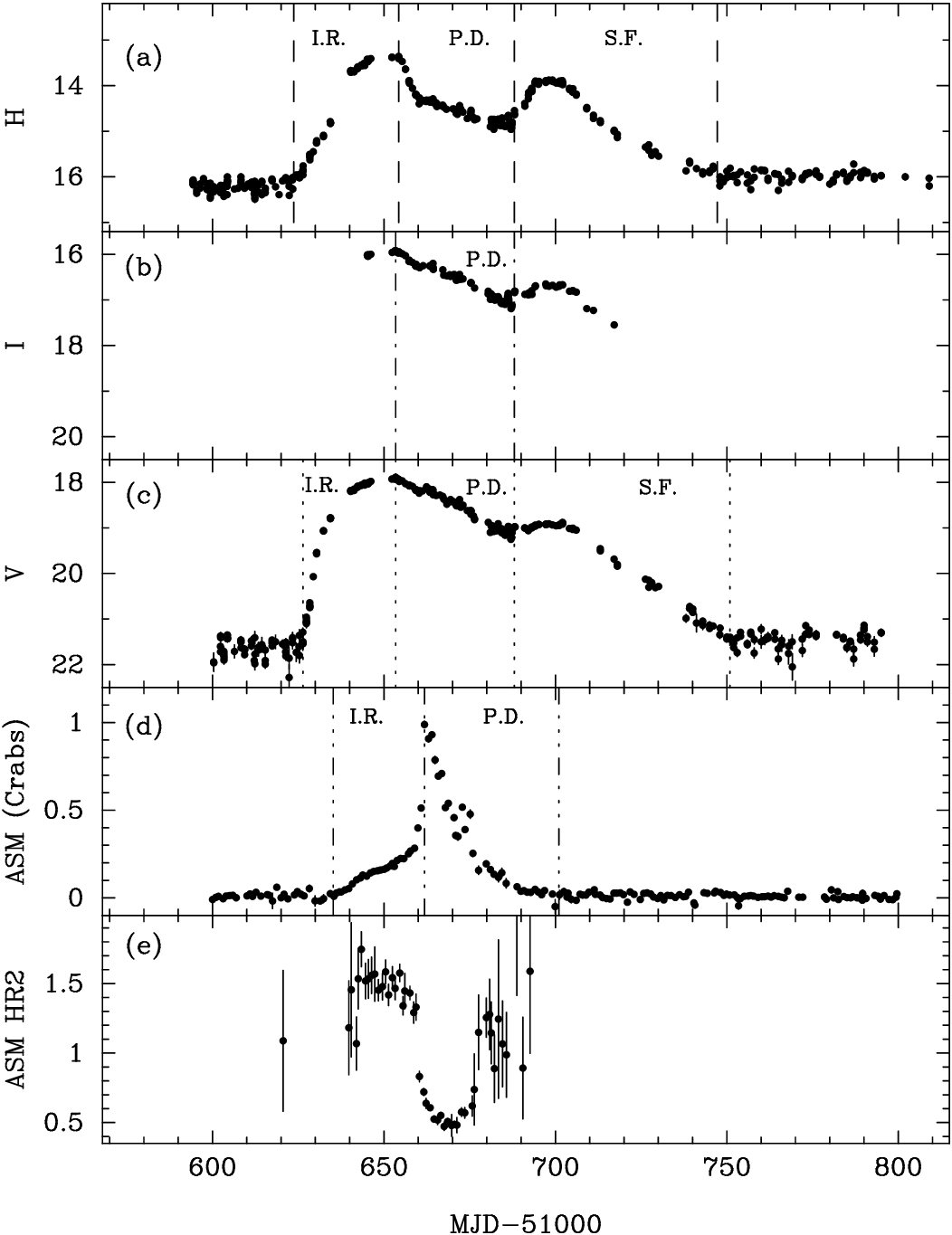]{ From top to bottom: YALO $H$, $I$, $V$, RXTE/ASM,
and ASM HR2 = (5-12 keV/3-5 keV) light curves.  The distinctive morphological
features, when appropriate, are denoted by the vertical lines, where I. R.,
P. D., and S. F., denote the initial rise, primary decay and secondary flare,
respectively (see Table 1 for exact dates and corresponding light curve
properties).  Note the prominent secondary flare, which is absent in the
X-rays, but coincides with a state transition apparent in the hardness ratio
curve.
\label{fig1}
}

\figcaption[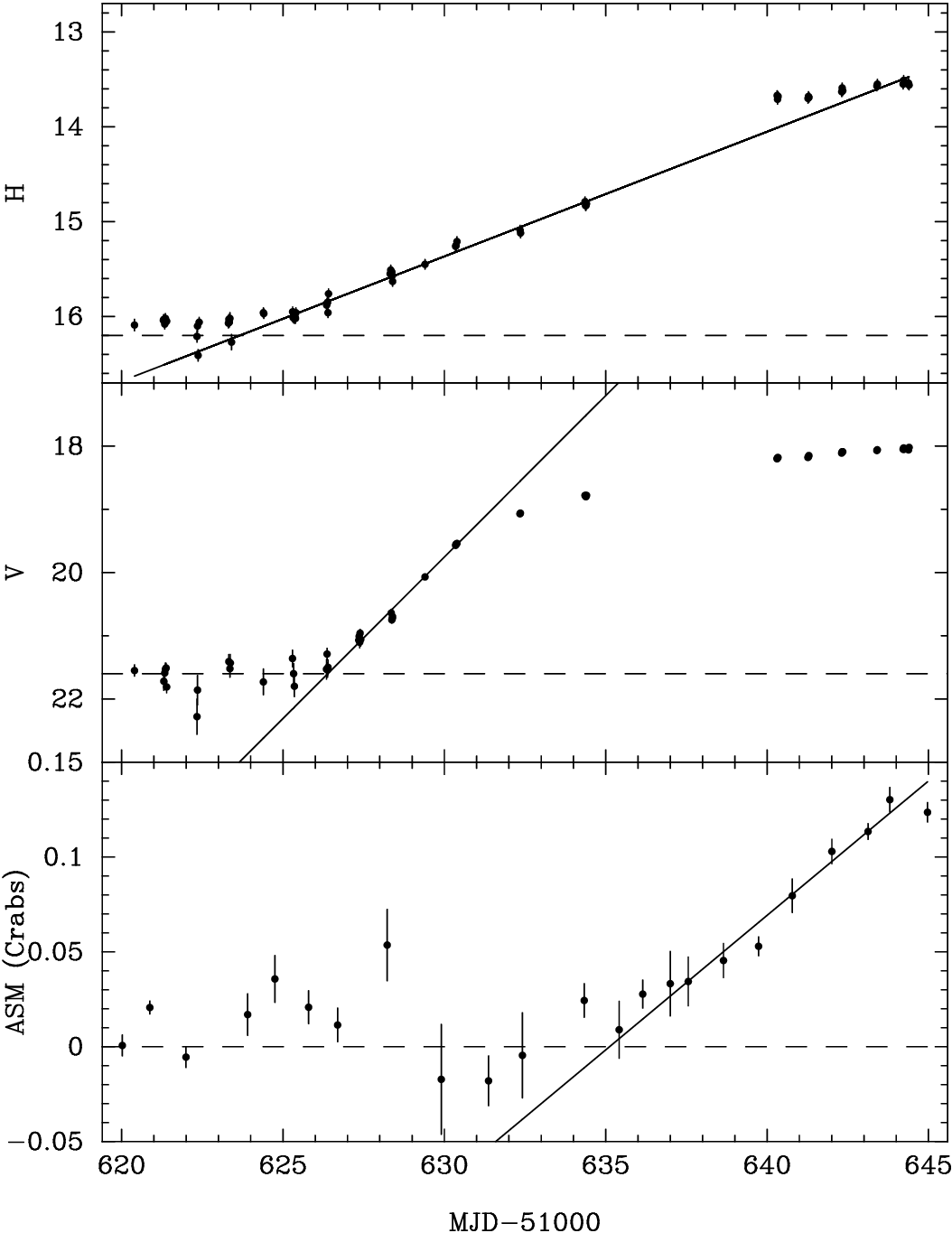]{Top to bottom: The $H$, $V$, and RXTE/ASM light curves
during the initial rise.  A linear fit was made to the data between MJD
51625--51635, MJD 51626--51631, and MJD 51635--51644 in $H$, $V$, and ASM,
respectively.  Based on these fits
the start of the  X-ray rise is delayed by of $8.8 \pm 0.6$ and $ 11.5 \pm
0.9$ days relative to the \v and \h band,  respectively.
\label{fig2}
}

\figcaption[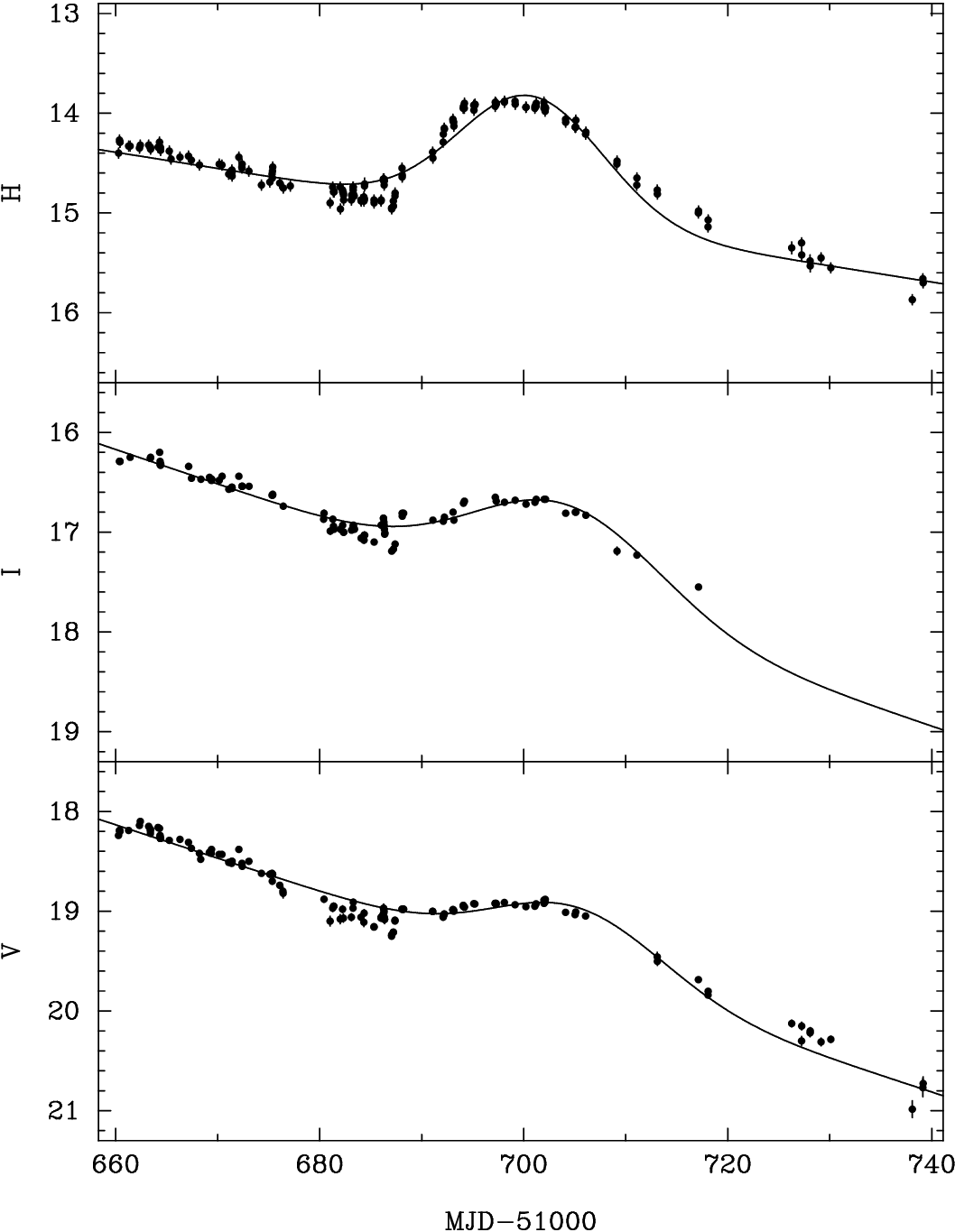]{Top to Bottom: A Gaussian plus linear fit to the
\vih data during the secondary flare (see Table~\ref{tab:lc}).  
quantifying the width and peak of the flare.  The best fit values to
the times, size, and width of the Gaussians are recorded in Table 1. 
The 
secondary peak is significantly sharper in $H$, compared to those of \i and \v
which occurred several days after the \h band peak.
\label{fig3}
}

\begin{figure}
\plotone{rkjfig1.ps}
\figurenum{1}
\caption{}
\end{figure}

\begin{figure}
\plotone{rkjfig2.ps}
\figurenum{2}
\caption{}
\end{figure}

\begin{figure}
\plotone{rkjfig3.ps}
\figurenum{3}
\caption{}
\end{figure}

\end{document}